\documentclass[aps,pre,twocolumn,superscriptaddress,showpacs,showkeys]{revtex4}
\usepackage{graphicx,bm,amssymb,amsmath,dcolumn,hyperref} 
\usepackage{enumerate}
\usepackage{float} 
\usepackage{wrapfig} 
\usepackage{amsfonts} 
\usepackage{slashbox}

\begin{document}

\title{Finite size scaling theory for percolation with multiple giant clusters}
\author{Yong Zhu}
\author{Xiaosong Chen}
\affiliation{State Key Laboratory of Theoretical Physics, Institute of Theoretical Physics, Chinese Academy of Sciences, P.O. Box 2735, Beijing 100190, China}

\date{\today}
\begin{abstract}
A approach of finite size scaling theory for discontinous percolation with multiple giant clusters is developed in this paper. The percolation in generalized Bohman-Frieze-Wormald (BFW) model has already been proved to be discontinuous phase transition. In the evolution process, the size of largest cluster $s_1$ increases in a stairscase way and its fluctuation shows a series of peaks corresponding to the jumps of $s_1$ from one stair to another. Several largest jumps of the size of largest cluster from single edge are studied by extensive Monte Carlo simulation. $\overline{\Delta}_k(N)$ which is the mean of the $k$th largest jump of largest cluster, $\overline{r}_k(N)$ which is the corresponding averaged edge density, $\sigma_{\Delta,k}(N)$ which is the standard deviation of $\Delta_k$ and $\sigma_{r,k}(N)$ which is the standard deviation of $r_k$ are analyzed. Rich power law behaviours are found for $\overline{r}_k(N)$, $\sigma_{\Delta,k}(N)$ and $\sigma_{r,k}(N)$  with critical exponents denoted as $1/\nu_1$, $(\beta/\nu)_2$ and $1/\nu_2$. Unlike continuous percolation where the exact critical thresholds and critical exponent $1/\nu_1$ are used for finite size scaling, the size-dependent pseudo critical thresholds $\overline{r}_k(N)$ and $1/\nu_2$ works for the data collapse of the curves of largest cluster and its fluctuation in discontinuous percolation in BFW model. Further, data collapse can be obtained part by part. That is, $s_1(r,N)$ can be collapsed for each jump from one stair to another and its fluctuation can be collapsed around each peak with the corresponding $\overline{r}_k(N)$ and $1/\nu_2$.

\end{abstract}
\maketitle

\section{Introduction}
Percolation transition represents the emergence of cluster large enough to compare with the system size as the density of occupied bonds or sites incerases. In classical percolation, bonds or sites are selected and occupied randomly and independently. It is continuous and  accompanied by universal critical exponents depending only on dimensionality of the system and  the number of components of the order parameter in both regular lattice of different dimensions and complex networks with different topologies.  Instead of occupation bonds or sites randomly, Achilioptas process proposed in Ref.\cite{Achlioptas} refers to adding one edge or site of some candidates on some certain conditions at each step and the percolation under product rule was claimed to be discontinuous in this seminal papaer. However, percolation transition under AP is identified to be continuous by both rigorous prove~\cite{Riordan11} and analysis of MC simulation data~\cite{Costa10,Liu11,Grassberger11,Fan12}. It is the different critical exponents that makes the explosive percolation unusual.  

Recently, a so called BFW model introduced by Bohman, Frieze and Wormald~\cite{Bohman} has attrated much attention~\cite{Chenw11,Chenw12,schrenk12,Zhangy12,Chenw13a,Chenw13b,Zhangrq13,Chenw14}. The BFW model starts N isolated nodes and a parameter $k$ which indicates the largest cluster allowed and is initialized as 2. At each step one edge is selected uniformly at random and it would accepted only if the size of the component formed from it is not larger than $k$. Otherwise $k$ would be augmented to $k+1$ and the edge would be rejudged if the fraction of accepted edges is smaller than a function $g(k)=1/2+(2k)^{-1/2}$ while a new edge would sampled randomly if the fraction of accepted edges is not. Asymptotically, half of the sampled edges would be accepted. This model is introduced to avoid the appearance of giant component.

Chen and D'Souza~\cite{Chenw11} generalized BFW model by taking the asymptotic fraction of accepted edges as a parameter $\alpha$ which is 0.5 in the original BFW model. The generalized BFW model exhibits percoaltion transition with the simultaneous emergence of multiple giant clusters which are stable and the transition is discontinuous for all $\alpha$ belongs to(0,0.97]. As $\alpha$ decreases, the number of giant clusters would increase, the sizes of giant clusters would decrease and the critical threshold of percolation transition would be delayed. When only one giant cluster exists and stops growing, a second giant component emerges in a continuous percolaiton transition whose critical points and exponents are dependent on $\alpha$. Through the analysis of BFW model generalized by taking $g(k)=0.5+(2k)^{\beta}$, they found the underlying mechanism for discontinuous transition is growth by overtaking~\cite{Chenw12}. BFW model on square and simple-cubic lattices was investigated by Schrenk and numerical evidence for strongly discontinuous transitions was found~\cite{schrenk12}. Zhang built a relation between $\alpha$ and the number of giant clusters~\cite{Zhangy12}. It indicated there would be only one giant cluster if $\alpha>0.52$. In Ref.\cite{Chenw13b}, the authors classified the BFM into two $\alpha$ regimes, one is the stable regime of a unique discontinuous transition where one or more giant components emerge and coexist throughout the supercritical regime while the other one is the unstable regime of multiple discontinuous transitions where multiple giant components emerge but the two smallest ones merge at a well-defined transition point in the supercritical regime.

Finite size scaling theory for continuous percolation transitions has been well established and has been used to characterize critical behavior of phase transition by critical threshold and critical exponents\cite{Liu11,Fan12,Privman84,Privman90}. For example, the reduced size of largest cluster of percolation on regular lattice follows a scaling form near the critical probability $p_c$
\begin{equation}
s_1=L^{-\beta/\nu} \, \tilde{s}_1((p-p_c)L^{1/\nu}). 
\end{equation}
where $L$ is the system's linear size, $\tilde{s}_1(z)$ is a scaling function, and $p$ is the occupation probability. $\beta$ and $\nu$ are the critical exponents associated with the size of largest cluster and the correlation length respectively. In the case of bond percolation on 2-dimensional square lattice, $p_c=0.5$, $\nu=4/3$ and $\beta=5/36$. In the case of percolation on random networks, $L$, $p$ and $p_c$ should be replaced by system size $N$, bond density $r$ (the ratio of the number of edges and the number of nodes in the graph) and critical threshold $r_c$ respectively.The critical exponents can be obtained with finite size scaling theory. It's know that $r_c=0.5$, $\nu=3$ and $\beta=1$ for percolation in ER random networks. Using finite size scaling theory, critical threshold and critical exponents can be determined with high precision. However, the finite size scaling theory for discontinuous percolation has not be established yet and it's a qustion that whether it exists. In this paper, we focus on the ffs behaviors of discontinuous transitions in BFW model. This paper is organised as follows: 

\section{Sampled quantities and finite size scaling theory for continuous percolation}
We do monte carlo simulations to study the generalized BFW($\alpha$) model with $\alpha \in [0.1,0.9]$ and $N$ from 10000 to 10240000. For each system size, we produced 512000 independent samples and used the algorithm of Newmann and Ziff~\cite{Newmann00,Newmann01} to track the size of largest cluster and second largest cluster. We denote $s_R$ as the reduced size of cluster ranked $R$ and $r$ as the edge density. In the process of adding edge one by one, the size of largest cluster ($s_1$) experiences a series of jump. The largest gap induced by single edge has been used to determine the continuity of percolation transition by its asymptotic behavior as $N\to \infty$. It converges to zero in power law for continuous percolation transitions while it converges to a nonzero value for discontinuous ones.

We sampled the following observables in our simulations:
 \begin{enumerate}[(1)]
 \item $s_1(r,N)$ and $s_2(r,N)$, size of largest and second largest cluster as a function of edge density $r$ and system size $N$.
 \item $\chi(r,N)\equiv N(<s_1^2>-<s_1>^2)$, the fluctuation of $s_1(r,N)$. In this paper, $\chi(r,N)$ is not defined as the standard deviation of size of largest cluster as in Ref.{} but in a similar form as the susceptibility per spin in Ising model, because it's more natural.
 \item $\overline{\Delta}_k(N)$ and $\overline{r}_k(N)$, the averaged $k$th largest gap of $s_1$ and mean of the corresponding edge density's. For example, $\overline{\Delta}_1(N)$ and $\overline{r}_1(N)$ represent the largest gap of $s_1$ induced by single edge and the edge density after adding that edge, and so on.
 \end{enumerate}
From these observables we calculated the following quantities:
 \begin{enumerate}[(i)]
 \item $\sigma_{\Delta,k}(N)\equiv \sqrt{<(\Delta_k-\overline{\Delta}_k(N))^2>}$, the standard deviation of $\Delta_k$.
 \item $\sigma_{r,k}(N)\equiv \sqrt{<(r_k-\overline{r}_k(N))^2>}$, the standard deviation of $r_k$.
 \end{enumerate}
$\overline{r}_1(N)$ can be regarded as the pseudo-critical threshold.

 For continuous percolation transition, $s_1(r,N)$ and $\chi(r,N)$ follow the finite size scaling form:
\begin{equation}
s_1(r,N)=N^{-\beta/\nu} \, \tilde{s}_1(tN^{1/\nu}). \label{scaling_s1}
\end{equation}
\begin{equation}
\chi(r,N)=N^{-\gamma/\nu} \, \tilde{\chi}(tN^{1/\nu}).\label{scaling_x}
\end{equation}
where $t=r-r_c$ characterizes the deviation from the critical point while $\beta$, $\gamma$ and $\nu$ are critical exponents. The ratio's of critical expoent satisfy the scaling relation:
\begin{equation}
2\beta/\nu+\gamma/\nu=1. \label{scaling_relation}
\end{equation}
Our previous works have shown that, for continuous percolation transition,
\begin{equation}
\overline{r}_k(N)-r_c(\infty) \propto N^{-1/\nu_1} 
\label{pl_r}
\end{equation}
where $r_c(\infty)$ is the real critical threshold.
\begin{equation}
\sigma_{r,k}(N) \propto N^{-1/\nu_2} 
\label{pl_sd_r}
\end{equation}
\begin{equation}
\overline{\Delta}_k(N) \propto N^{-(\beta/\nu)_1} 
\label{pl_delta}
\end{equation}
\begin{equation}
\sigma_{\Delta,k}(N) \propto N^{-(\beta/\nu)_2} 
\label{pl_sd_delta}
\end{equation}
In most instances, the exponents $1/\nu_1$ and $1/\nu_2$ are approximately equal to $1/\nu$ defined in Eq.\ref{scaling_s1} and \ref{scaling_x}, $(\beta/\nu)_1$ and $(\beta/\nu)_2$ are approximately equal to $\beta/\nu$ defined in Eq.\ref{scaling_s1}. However, in some instances such explosive percolation, $1/\nu=1/\nu_2\not=1/\nu_1$. So, $1/\nu_2$ and $(\beta/\nu)_1$ (or $(\beta/\nu)_2$) can be used to determine critical exponent and get the finite size scaling function of $s_1(r,N)$.

However, for discontinuous percolation transition, what we already know is that $\overline{\Delta}_k(N)$ converges to a nonzero constant in the thermodynamic limit $N\to \infty$. It can be inferred that $\beta/\nu=0$. But the asymptotic behaviors of $\overline{r}_k(N)$, $\sigma_{\Delta,k}(N)$, $\sigma_{r,k}(N)$ and the scaling form of $s_1(r,N)$ and $\chi(r,N)$ for discontinuous percolation transition have not been discussed so far.

\section{Order parameter and fluctuation}

\begin{figure*}
\includegraphics{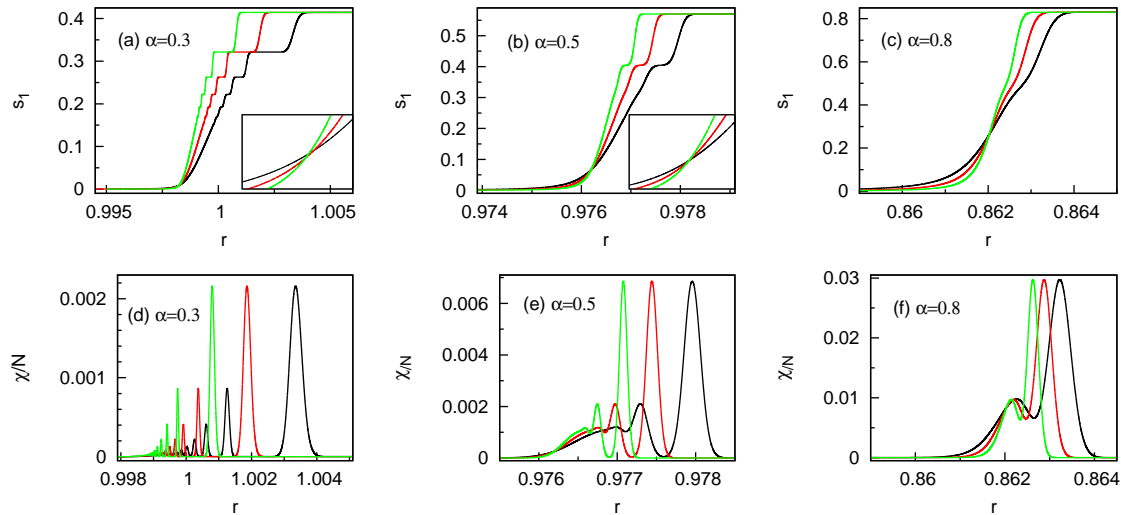}
\caption{Order parameter and its fluctuation in BFW model. (a) For $\alpha=0.3$, $s_1(r,N)$ versus $r$. (b) For $\alpha=0.5$, $s_1(r,N)$ versus $r$. (c) For $\alpha=0.8$, $s_1(r,N)$ versus $r$. (d) For $\alpha=0.3$, $\chi(r,N)$ versus $r$. (e) For $\alpha=0.5$, $\chi(r,N)$ versus $r$. (f) For $\alpha=0.8$, $\chi(r,N)$ versus $r$. Each plot includes three curves with system sizes $N=$2560000, 5120000, 10240000. The inserts in (a) and (b) show the behavior near the critical thresholds. Thresholds can be estimated from the intersections where $r$=0.99823(3), 0.97619(1) and 0.86202(1) for $\alpha=0.3, 0.5, 0.8$ respectively. Scaling of $s_1(r,N)$ and $\chi(r,N)$ are shown in Fig.\ref{fig3}.}
\label{fig1}
\end{figure*}

In the study of percolation transition, the size of largest cluster $s_1(r,N)$ plays the role of parameter while its fluctuation $\chi(r,N)$ plays the role similar to susceptibility per spin in Ising model. In case of continuous percolation, $s_1(r,N)$ increases conitnuously and $\chi$ diverges at critical threshold. The dievergence of $\chi$ is reflected by a peak for finite size and the edge density $r$ where $\chi(r,N)$ shows a peak converges to the critical threshold in power law as $N\to \infty$.

In Fig.\ref{fig1}, Curves of $s_1(r,N)$ and $\chi/N$ are plotted with respect to $r$ for $\alpha=0.3,0.5,0.8$ with three different system sizes $N=$2560000,5120000,10240000. 

When $\alpha=0.3$ shown in Fig.\ref{fig1}~(a) and (d), in the supercritical regime where the percolation has already took place, $s_1(r,N)$ does not increase continuously but in a staircase way and $\chi/N$ shows a series of peaks instead of only one peak. At least six stairs and six peaks can be found by simply zooming in the plots and every peak of $\chi/N$ corresponds to a jump of $s_1(r,N)$ from one stair to another. Both of the heights of stairs in curves of $s_1(r,N)$ and peaks in curves of $\chi/N$ are independent of system size. The heights of stairs from up to bottom in curves of $s_1(r,N)$ are about 0.4143, 0.3212, 0.2624, 0.2223, 0.1924, 0.1697, 0.1517, and so on. Numerically, $0.4143\approx0.2223+0.1924$ and $0.3212\approx0.1697+0.1517$. This reflects the mechanism of percolation transition in BFW model which is growth by overtaking. To be specific, in the evolution process of BFW model, two smaller giant cluster combine and form a new giant cluster which is larger than the previous largest cluster. Then the new cluster becomes the largest one and the previous largest one becomes the second largest. In Ref.{}, growth by overtaking is argued to be the reason discontinuous percolation transition. Also, we can infer that three giant clusters survives at last since clusters whose sizes are smaller than the third stairs from above are merged, which agrees with the results in Ref.{}.

The insert of Fig.\ref{fig1}(a) demonstrates the fixed point of $s_1(r,N)$ with different system sizes at $r=0.9982$. In the case of continuous percolation, the intersection of $s_2/s_1$ with different system sizes has been used to estimate the critical threshold since accoding to finite size scaling theory,
\begin{equation}
s_2/s_1=\tilde{s}_2(tL^{1/\nu}) /\tilde{s}_1(tL^{1/\nu})\equiv
U(tL^{1/\nu}).  
\end{equation}
Obviously, the ratio at $t=0$ is independent of the system size $L$. But in the case of discontinuous percolation, if Eq.\ref{scaling_s1} still holds and $\beta/\nu$ is substituted by 0, then curves of $s_1(r,N)$ with different system sizes are expected to intersect at the critical threshold of percolation transition. Also, if Eq.\ref{scaling_x} and \ref{scaling_relation} still hold, then we would get $\gamma/\nu=1$ and $\chi(r,N)\propto N$ for each level of peaks which can be inferred from Fig.\ref{fig1}~(d).

From other subfigures in Fig.\ref{fig1}, we can see that the scenario discussed above for $\alpha=0.3$ holds for $\alpha$ with other values. Since $\alpha$ reprensents the asymptotical acceptance rate of sampled edges, the suppression of $s_1(r,N)$ is stronger with smaller $\alpha$. From the horizontal comparison, we can find that as $\alpha$ increases, the numbers of stairs stairs in curves of $s_1(r,N)$ and peaks in curves of $\chi/N$ decrease, the final size of $s_1(r,N)$ increases and the critical threshold becomes smaller. The critical thresholds estimated from the intersection of $s_1(r,N)$ are 0.9762 and 0.862 for $\alpha=0.5,~0.8$ respectively. All the results are summarized as $r_c^1$ in Table.\ref{table1}.

\section{Analysis of Critical thresholds and critical exponent}
\begin{table*}
\caption{Summary of critical thersholds and critical exponents. $r_c^1$ is obtained from intersection of $r_1(r,N)$ with different system sizes and $r_c^2$ is obtained from the convegence of $r_k(N)$. $(\beta/\nu)_2$ is obtained from $ln(\sigma_{\Delta}^k(N))$,$1/\nu_1$ is obtained from $ln(r_k(N)-r_c)$ and $1/\nu_2$ is obtained from $ln(\sigma_{r}^k(N))$. There are three columns for $1/\nu_2$ because $1/\nu_2$ for the three largest jump is not always the same.}
\begin{ruledtabular}
\begin{tabular}{c|c|c|c|c|ccc}
$\alpha$ & $r_c^1$ & $r_c^2$ & $(\beta/\nu)_2$ & $1/\nu_1$ & \multicolumn{3}{c}{$1/\nu_2$} \\ \hline
0.1 & 1.0003(2) & 0.9995(10) & 0.39(2) & 0.45(3) & 0.687(38) & 0.96(28) &        \\
0.2 & 1.00003(6) & 1.0000(5) & 0.39(2) & 0.489(30) & 0.74(1) & 0.75(4) & 0.80(15) \\
0.3 & 0.99823(3) & 0.99821(2) & 0.39(1) & 0.49(2) & 0.738(2) & 0.72(2) & 0.70(4)  \\
0.4 & 0.99119(2) & 0.9912(2) & 0.38(1) & 0.495(20) & 0.720(2) & 0.58(4) & 0.55(8) \\
0.5 & 0.97619(1) & 0.9762(1) & 0.38(1) & 0.498(16) & 0.553(30) & 0.520(5) & 0.522(10) \\
0.6 & 0.95151(1) & 0.9515(1) & 0.38(1) & 0.489(20) & 0.704(2) & 0.512(12) & 0.507(13)  \\
0.7 & 0.91499(1) & 0.91499(4) & 0.38(1) & 0.492(20) & 0.537(21) & 0.505(6) & 0.504(11) \\
0.8 & 0.86202(1) & 0.86200(4) & 0.37(1) & 0.495(20) & 0.509(7) & 0.503(5) & 0.504(9) \\
0.9 & 0.77994(1) & 0.77994(4) & 0.37(1) & 0.495(10) & 0.503(5) & 0.502(7) & 0.505(13) \\
\end{tabular}
\end{ruledtabular}
\label{table1}
\end{table*}

In this section, we will check whether power law relations in Eq.\ref{pl_r}-\ref{pl_sd_delta} hold or not for discontinuous percolation transition in BFW model. Three largest jumps in $s_1(r,N)$ from the addition of a single edge are discussed here. 

\begin{figure}
\resizebox{0.45\textwidth}{0.32\textwidth}{ \includegraphics{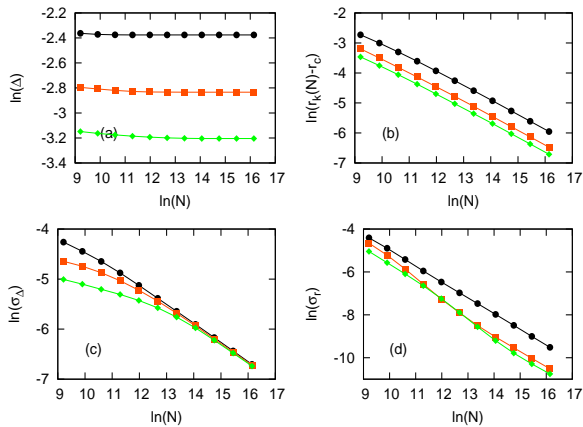} }
\caption{Plots of $ln(\Delta_k)$, $ln(r_k(N)-r_c)$, $ln(\sigma_{\Delta}^k(N))$ and $ln(\sigma_{r}^k(N))$ versus $ln(N)$ for $\alpha=0.3$. The $r_c$ used in (b) is 0.9982.}
\label{fig2}
\end{figure}

Fig.\ref{fig2} shows the details when $\alpha=0.3$: not only the largest jump but also the second and third (even more but not shown) largest jump of $s_1(r,N)$ converge to a positive constant. This is strong evidence for discontinuous percolation. $r_k(N)$ converges in power law to a same value which is 0.9982 and equal to the critical threshold estimated from the intersection of $s_1(r,N)$ with different system sizes. Both $\sigma_{\Delta}^k(N)$ and $\sigma_{r}^k(N)$ converge to zero in power law. Moreover, within error bar, critical exponent $1/\nu_1$ caculated from the slopes of three curves of $ln(r_k(N)-r_c)$ versus $ln N$ in Fig.\ref{fig2}(b) is approximately 0.49, $(\beta/\nu)_2$ caculated from the slopes of three curves of $ln(\sigma_{\Delta}^k(N))$ versus $ln N$ in Fig.\ref{fig2}(c) is approximately 0.39 and $1/\nu_2$ caculated from the slopes of three curves of $ln(\sigma_{r}^k(N))$ versus $ln N$ in Fig.\ref{fig2}(d) is approximately 0.74, 0.72 and 0.70. Data of small $N$ are discarded in order to get more accurate slope.

Power law behaviors holds for other values of $\alpha$. All the results of critical thresholds and ciritcal exponents are summarized in Table.\ref{table1}. The critical thresholds estimated from the intersection of $s_1(r,N)$ and from the convegence of $r_k(N)$ are always the same. As $\alpha$ increases, $(\beta/\nu)_2$ remains the same although there's slight decreasing from 0.39(2) to 0.37(1). $1/\nu_1$ is always the same in the three largest jump with different $\alpha$. Except when $\alpha=0.1$, $1/\nu_1$ seems to be equal to 0.5. However the exponent $1/\nu_2$ is not always the same for three largest jump. Furthermore, $1/\nu_2$ decreases to 0.5 so as to equal to $1/\nu_1$. As we know, BFW model leads to ER model when $\alpha=1$. It's believed there's a crossover regime where the discontinuous percolation becomes continuous when $\alpha$ is large enough and it deserves further investigation.

\section{Finite size scaling theory for discontinuous percolation}
\begin{figure*}
\includegraphics{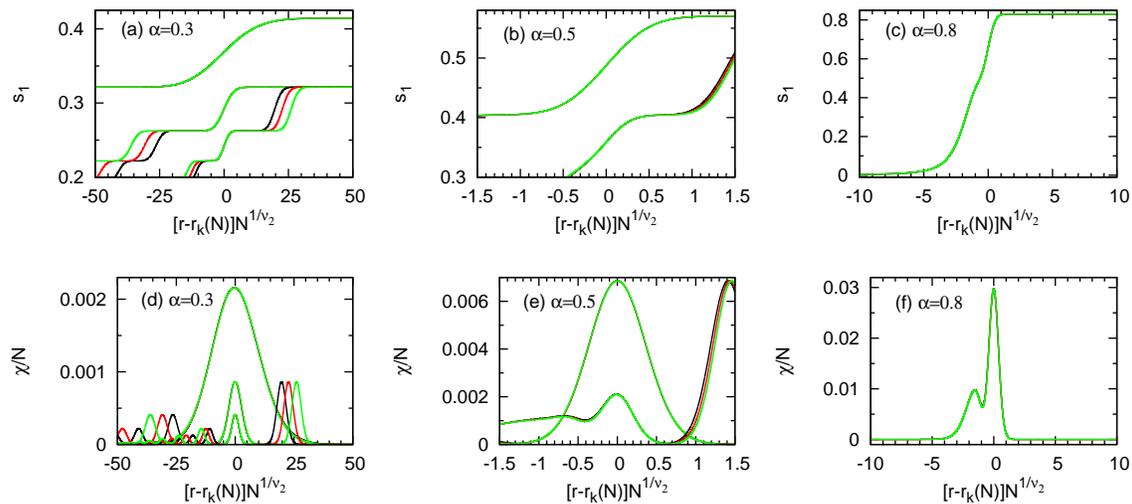}
\caption{Data collapse of $s_1(r,N)$ and $\chi/N$ in BFW model for $\alpha=0.3, 0.5, 0.8$. (a) For $\alpha=0.3$, curves of $s_1(r,N)$ in the range of three largest jump from one stair to another are collapsed together with the corresponding $r_k(N)$ and $1/\nu_2$ respectively. (b) For $\alpha=0.5$, curves of $s_1(r,N)$ in the range of two largest jump from one stair to another are collapsed together with the corresponding $r_k(N)$ and $1/\nu_2$ respectively (c) For $\alpha=0.8$, curves of $s_1(r,N)$ are collapsed together with $r_1(N)$ and $1/\nu_2$. (d) For $\alpha=0.3$, curves of $\chi/N$ around the three highest peaks are collapsed together with the corresponding $r_k(N)$ and $1/\nu_2$ respectively. (e) For $\alpha=0.5$, curves of $\chi/N$ around the two highest peaks are collapsed together with the corresponding $r_k(N)$ and $1/\nu_2$ respectively. (f) For $\alpha=0.8$, curves of $\chi/N$ are collapsed together with $r_1(N)$ and $1/\nu_2$. Data with system sizes $N=$2560000, 5120000, 10240000 are used.}
\label{fig3}
\end{figure*}

For continuous percolation, what if $t=r-r_c$ is replaced by $t^{'}=r-r_k(N)$~?
\begin{equation}
\begin{aligned}
t^{'}L^{1/\nu}&=\left(r-r_k(N)\right)L^{1/\nu} \\
		  &=\left(r-\left(r_c+const.~L^{-1/\nu_1}\right)\right)L^{1/\nu} \\
		  &=\left(r-r_c\right)L^{1/\nu}-const.~L^{1/\nu-1/\nu_1}	\\
		  &=tL^{1/\nu}-const.
\end{aligned}
\label{new_t}
\end{equation}
The convergence of $r_k(N)$ in power law has been applied in the second line and $\beta/\nu=(\beta/\nu)_1$ has been applied in the fourth line in Eq.\ref{new_t}. Obviously, the only impact is a constant horizontal shift, which is independent of $N$, of scaling function $\tilde{s}_1(z)$ and $\tilde{\chi}(z)$.

The situation for discontinuous percolation is much more different. In the cases where $1/\nu_1=1\nu_2$, curves of $s_1(r,N)$ or $\chi/N$ with different system sizes can indeed be collapsed together by the finite size scaling form in Eq.\ref{scaling_s1} and \ref{scaling_x} with $1/\nu_1$ or $1/\nu_2$. In Fig.\ref{fig3}(c), data of $s_1(r,N)$ for $\alpha=0.8$ are perfectly collapsed together although there seems to be a contraflexure point. In Fig.\ref{fig3}(f), data of $\chi/N$ for $\alpha=0.8$ are also perfectly collapsed together although there are two peaks. As proved above, both $r_c$ and $r_k(N)$ can be used and $r_1(N)$ has been used in (c) and (f). However, in the cases where $1/\nu_1\not=1\nu_2$, it failed when $1/\nu_1$ is used with both $r_c$ and $r_k(N)$. So, it's possible that it is $1/\nu_2$ instead of $1/\nu_1$ that works for $\alpha=0.8, 0.9$. It turns out to hit the truth. 

Data collpase of $s_1(r,N)$ and $\chi/N$ for $\alpha=0.3$ and $\alpha=0.5$ are shown in other subfigures in Fig.\ref{fig3}. Curves of $s_1(r,N)$ and $\chi/N$ can only be collapsed together piecewise. To be specific, $s_1(r,N)$ can be scaled for each jump from one stair to another and $\chi/N$ can be scaled around each peak with the corresponding $r_k(N)$ and $1/\nu_2$. When $\alpha=0.3$ shown in (a) and (d), three parts of $s_1(r,N)$ and three peaks of $\chi/N$ are collapsed together with $r_1(N)$ and $1/\nu_2=0.738$, $r_2(N)$ and $1/\nu_2=0.72$, $r_3(N)$ and $1/\nu_2=0.70$, respectively. When $\alpha=0.5$ shown in (a) and (d), two parts of $s_1(r,N)$ and two peaks of $\chi/N$ are collapsed together with $r_1(N)$ and $1/\nu_2=0.553$, $r_2(N)$ and $1/\nu_2=0.520$, respectively. 

In the Ref.\cite{Chenw13b}, the authors clarify generalized BFW model into two regimes. All the cases discussed above belong to the stable regime of a unique discontinuous percolation. Our approach also works well for the unstable regime (see Appendix).

\section{Summary}
We have investigated discontinuous percolation in generalized BFW model and developed a approach of finite size scaling for it. We analysis the behaviors of the $k$th largest jump, induced by single edge, of the size of largest cluster with $k$ up to 3. Similar to continuous percolation, rich power laws are found. The mean $k$th largest jump $\Delta_k(N)$ converges to a positive constant indicating the discontinuity of percolation. The corresponding averaged edge densities $r_k(N)$ always converge to the real critical thresholds in power law with exponents $1/\nu_1$ which is 0.5 except when $\alpha=0.1$. The standard deviation of $\Delta_k(N)$ converges to zero in power law with exponent $(\beta/\nu)_2$ which differs slightly for different $\alpha$. The standard deviation of $r_k(N)$ converges to zero in power law with exponent $1/\nu_2$ which is not always the same for the three largest jump while it is always the same for continuous percolation. $1/\nu_2$ converges to 0.5 as $\alpha$ increases.

For continuous percolation, the real critical thresholds and critical exponents $1/\nu_1$ take the responsibility of finite size scaling with the scaling variable $(r-r_c)L^{1/\nu_1}$. However, for discontinuous percolation in BFW model, data collapse can be achieved with the size-dependent pseudo critical thresholds $r_k(N)$ and critcal exponents $1/\nu_2$ from the corresponding standard deviation of $r_k(N)$. When $\alpha=0.8, 0.9$, the critical exponents $1/\nu_2$ from three largest jumps are approximately the same and data collapse can be done once and for all. For the cases with different $1/\nu_2$ from three largest jumps, data collapse can be attained piecewise. The part of $s_1$ where the largest jump takes place and the part of $\chi/N$ around the highest peak can be collapsed together with $r_1(N)$ and $1/\nu_2$ from $\sigma_r^1(N)$, the part of $s_1$ where the second largest jump takes place and the part of $\chi/N$ around the second highest peak can be collapsed together with $r_2(N)$ and $1/\nu_2$ from $\sigma_r^2(N)$, and so on.

\begin{acknowledgments}

\end{acknowledgments}

\end{document}